\documentclass{article}

\usepackage[utf8]{inputenc}
\usepackage[T1]{fontenc}
\usepackage{graphicx}
\usepackage{hyperref}
\usepackage{listings}
\usepackage{url}

\newcommand\ipaddr[4]{\texttt{#1.#2.#3.#4}}
\newcommand\localhost{\ipaddr{127}{$\ast$}{$\ast$}{$\ast$}}
\newcommand\hfield[1]{``\texttt{#1}''}
\newcommand\attrname[1]{``\texttt{#1}''}
\newcommand\attrvalue[1]{``\texttt{#1}''}

\begin{document}

\title{An Architecture to Support the Invocation of Personal Services in Web Interactions}

\author{André Zúquete, Fábio Marques\\
\{andre.zuquete,fabio\}@ua.pt}

\maketitle

\begin{abstract}
This paper proposes an architecture to enable Web service providers to
interact with personal services. Personal services are vanilla HTTP
services that are invoked from a browser, upon a request made by a
service Provider, to deliver some service on the client side, i.e.,
on an execution environment defined by the browser's user. Personal
services can be used both to handle content manipulation and
presentation or to deliver request-response interactions with
different goals (e.g. user authentication).  Unlike plugins, that
are described to service providers on each and every HTTP request,
personal services are explicitly searched by service providers using
a novel agent, a Broker, that works in close cooperation with each
browser. We have implemented this architecture and implemented an
HTTP proxy to cope with it.  For demonstration purposes we show how
we can use personal services for
personal authentication
with an electronic identification (eID) card.
\end{abstract}

\newcommand\keywords[1]{\textbf{Keywords}: #1}

\section{Introduction}

Web-based interactions have become one fundamental part of Internet
interactions, and browsers are the most frequent applications used
as clients in those interactions. Web browsers were initially
conceived as information presenters and navigators, but
progressively became the omnipresent graphical interface of complex,
Web-based applications.

Web-based applications often benefit from (or require) special
handling services provided by client browsers. Plugins were invented
for this, e.g. for handling special kinds of data provided in Web
servers' responses. But plugins are a nightmare to manage, they are
browser-specific and they need to follow browsers' updates.

In this paper we present a new paradigm to extend Web applications
by using the concept of \textit{personal services}. Personal services are
vanilla HTTP services that are explored from a browser to the benefit
of its user. Personal services may run locally to browsers or not.

\subsection{Personal services}

To clarify the text, we will use the terms \textit{browser},
\textit{service provider} (SP) and \textit{personal service}.  The
terms client and server will be avoided because they may be used in
many ways: browsers are clients of SPs, SPs may be clients of
personal services and these will be servers of both browsers and SPs.

Personal services are technically HTTP services. These services can
be invoked by other services, namely SPs, in order to provide some
service to (or from) a user. A personal service can either provide
some kind of content handling task (e.g. present a media object) or
can tackle request-response interactions (e.g. eID-based
authentications).

The HTTP paradigm is based on client-server, unilateral
request-response dialogues. The most common HTTP clients, the
browsers, make a request to a server, collect its response and
process it. In this paradigm it is not normal to have servers to
make direct requests to other servers; instead, servers invoke other
servers through a client redirection. We decided to kept this
paradigm for dealing with personal services, in order to facilitate
the adaptation of browsers to handle them. Namely, we enriched the
HTTP redirection mechanism to fully handle the exploitation of
personal services from SPs.

\subsection{HTTP redirections}
\label{http.redirections}

HTTP redirections enable an HTTP server to redirect a client request
to an alternative URL (Uniform Resource Locator). The redirection
may be performed in many ways, for dealing with different
redirection reasons, using a specific HTTP status code of the $3$xx
class~\cite{rfc7231}.

\begin{figure}[ht]
\centerline{
\includegraphics[scale=.55]{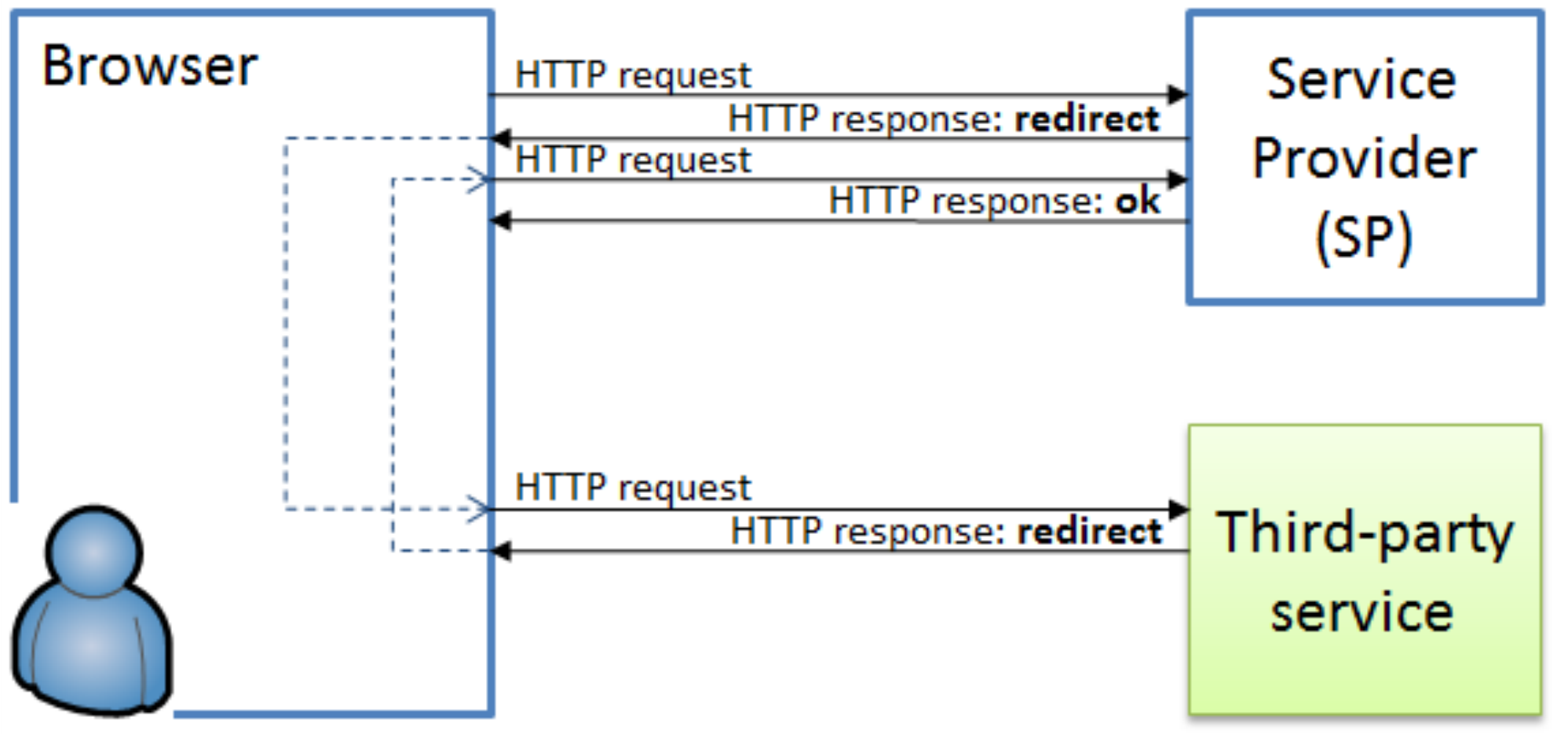}
}
\caption{Invocation of third-party services from a browser upon a
redirection response returned by an SP.}
\label{Redirect.Fig}
\end{figure}

Although HTTP redirections were initially conceived for dealing
with temporary or permanent relocations of SPs, currently they are
also used to handle indirect calls of SPs to third-party services mediated by a
browser (see Figure~\ref{Redirect.Fig}). An example of this is the
OASIS Web Browser SSO Profile~\cite{SAMLProfiles}, using SAML
(Security Assertion Markup Language), which is used to identify and
authenticate a browser user with a central, third-party service
acting as an Identity Provider (IdP).

The problem of using HTTP redirections to invoke personal services
is that their exact location, in terms of TCP/IP address, are
unknown to SPs; that depends of the client. Unlike with central,
third-party services explored by SPs, which know their TCP/IP
address, the location of personal services must be somehow provided
by each browser on a case-by-case basis.

In~\cite{Zuquete14} we explored the concept of personal services to
extent the OASIS SSO authentication process with a Personal IdP
(PIdP).  This PIdP was able to interact with a local eID token in
order to perform a user identification and authentication with it.
However, we had to use a fixed IP address (\ipaddr{127}{0}{0}{1}) and a fixed
port (666) to reach the PIdP, which was not a clean and flexible
solution. In this paper we propose a solution to deal with this
problem by allowing an SP to discover the presence of the intended
PIdP (which is a personal service) and to properly invoke it.

\subsection{Contribution}

In this paper we propose a solution for enabling SPs to discover
useful personal services and to invoke them upon such discovery. To
do so, we conceived:
\begin{enumerate}
\item A name system (Broker) that can be used by SPs to discover a
      personal service of interest;
\item A name resolution protocol, which provides a handle for
      invoking a particular personal service given its name; and
\item A personal service invocation protocol, using
      previously obtained handles.
\end{enumerate}

The Broker is a service running locally to a browser that keeps track
of all personal services available to the browser. It is not
a browser extension or plugin, it is a separate service that can be
used by any browser a user chooses to use. The Broker uses
attribute-based descriptions of personal services to enable SPs to
look for services of interest, similarly to the Yellow Pages service
described in~\cite{Peterson87} and to the UDDI (Universal
Description, Discovery \& Integration) Web Services' discovery
paradigm~\cite{UDDI04}. Attribute-based descriptions are, in our
opinion, flexible enough to handle the description of the
functionalities of personal services that may be looked by SPs.

The Broker is responsible for instantiating all personal services, that are locally hosted to the browser,
prior to their actual invocation by SPs. Thus, local personal services do
not need to stay running all the time, they can be executed on
demand, and just in time, to handle SP invocations.
Remote personal services are not controller by the Broker, so it cannot instantiate them. Instead, the Broker, is responsible to verify if the remote personal services are active.

\begin{figure}[p]
\centerline{
\includegraphics[scale=.55]{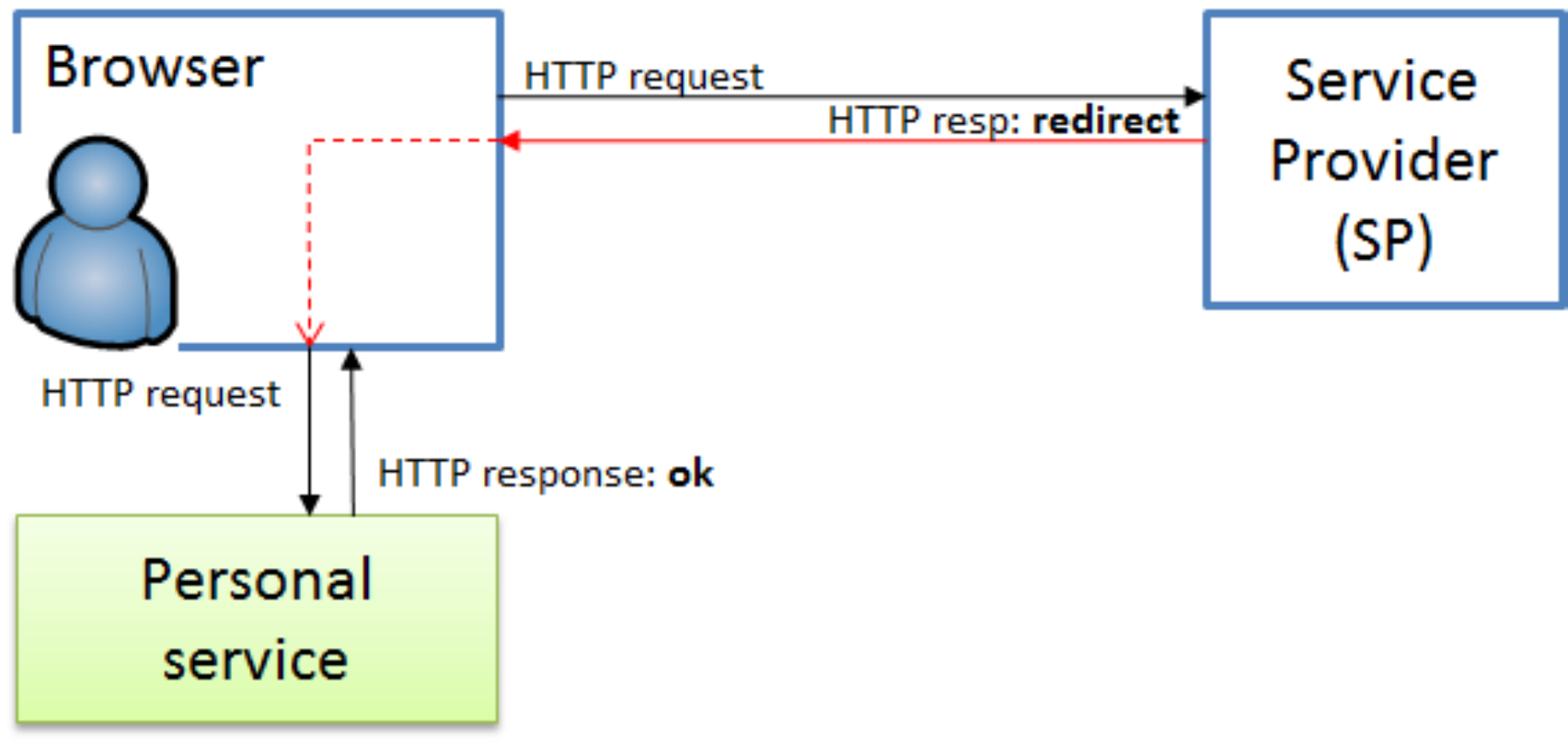}
}
\centerline{
\includegraphics[scale=.55]{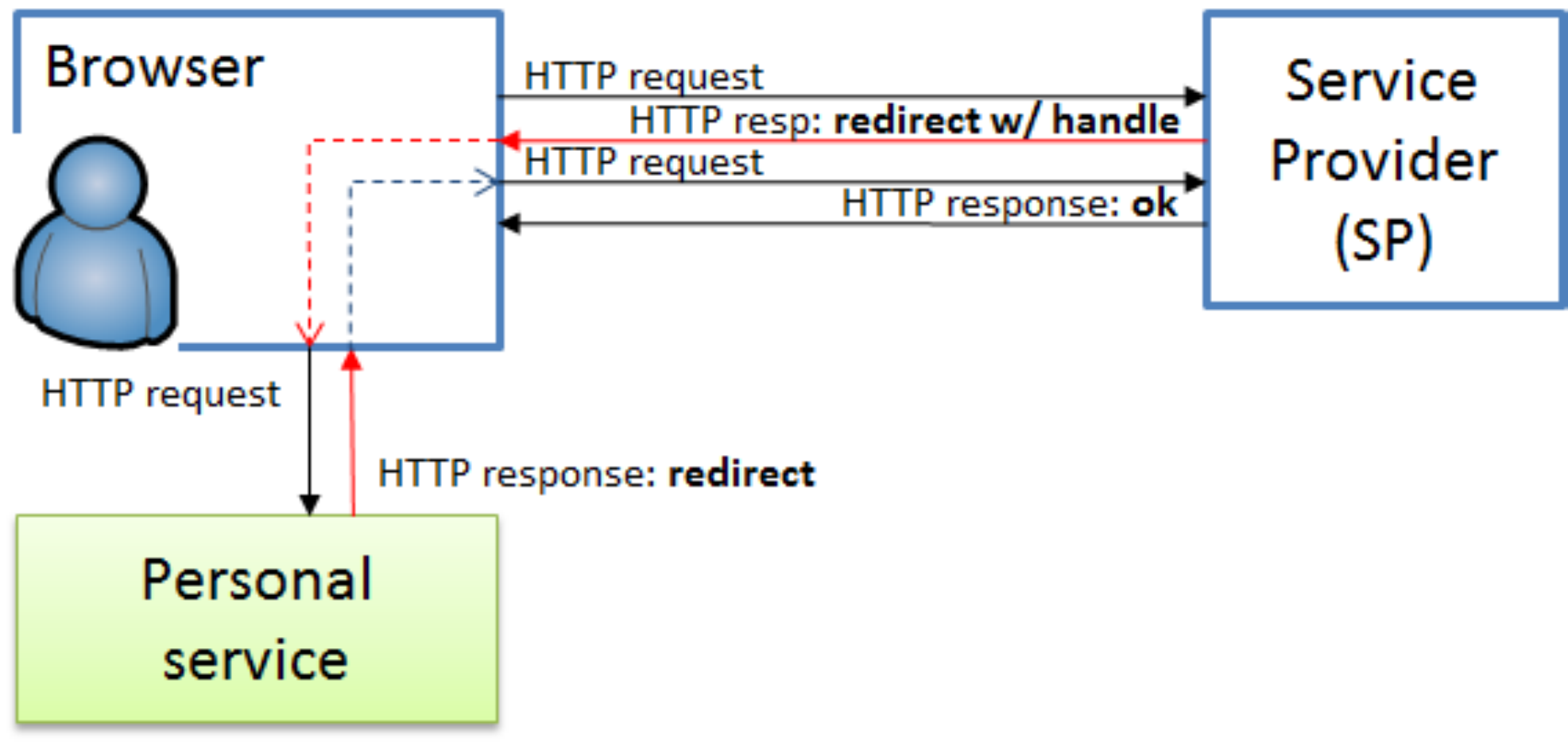}
}
\caption{Invocation of personal services from a browser upon a
redirection response with a service handle returned by an SP. The
personal service can terminate the HTTP request-response dialog (top
diagram) or redirect it again to the SP, or even to another SP
(bottom diagram). This last redirection is usually based on an URL
provided by the SP that indirectly invoked the personal service.}
\label{PS.Fig}
\end{figure}

Once activated, a personal service can be indirectly invoked from an
SP through a browser using a new HTTP redirection mechanism (see
Figure~\ref{PS.Fig}). The SP invocation is handle-based, i.e., the
SP refers to the personal service using a previously fetched handle.
The browser translates the handle into a local TCP/IP address,
which is then used in an HTTP request to the personal service using
data provided by the SP in the redirection reply. The personal
service can reply to the SP, if required, using the existing HTTP
redirection mechanism, very much like actual third-party reply to
the SPs that invoked them (see Figure~\ref{Redirect.Fig}). However,
personal services can also directly invoke their calling SP, or
other SPs, if conceived to do so; that is not forbidden by our proposal.

The handle-based redirection to personal services has several
advantages over a direct redirection using a TCP/IP address. Since
it adds a layer of abstraction to the service invocation, forcing
browsers to resolve the handle prior to invoke the personal service,
the exact TCP/IP address of personal services remains hidden from
SPs. This allows personal services to use variable addresses along
time without disrupting their call from handles cached by SPs. It
also allows SPs to get valid handles without requiring personal
services to be running before being actually invoked or,
alternatively, to be forced to use a fixed TCP/IP address. Finally,
since personal services do not need to be running before being
actually invoked, they cannot be discovered by SPs without the help
of the Broker (e.g., by doing exhaustive redirections to several TCP
ports bound to localhost addresses).

This paper is structured as follows.
Section~\ref{related.work} presents the related work.
Section~\ref{architecture} presents the architecture of our
contribution.
Section~\ref{implementation} presents the implementation of a
prototype that was developed for experimenting our contribution.
Section~\ref{experimentation} presents an experimentation of our
proposal with a personal service conceived for an eID-based personal
authentication, presented in~\cite{Zuquete14}.
Section~\ref{javascript} discusses the usage of the proposed
architecture by active code running on Web resources loaded by
browsers, namely JavaScript.
Finally, Section~\ref{conclusions} concludes the paper.

\section{Related work}
\label{related.work}

At first, browsers were mainly used for surfing the world-wide web,
but now they are being used to access and execute more complex and
complete Web applications, ranging from social media Web
applications to productivity suits. Consequently, browsers'
manufacturers started to allow users to install plugins and
extensions in order to expand the browsers' capabilities, allowing
users to have better experiences.

Plugins are applications that are executed from a browser and use a
fixed and limited API (Application Programming Interface) it
provides to access its own contents and user interface. They were
created mainly to handle specific content types (e.g. PDF documents)
and browsers announce their availability to service providers (on
HTTP request header fields).  But, besides the increase of
functionalities, plugins also brought some security issues (such as
the identification of users from their browsers'
profile~\cite{Eckersley10}), as well as maintenance and universality
problems (they need to keep up with the evolution of browsers and
they are not provided for all browsers).

Grier \textit{et al.}~\cite{Grier09} proposed to use separate
processes for keeping plugins away from browsers' address space. This
is similar to our proposal, but our goal is different. In fact, we
want to enable SPs to interact (via HTTP requests and
responses) with local, personal services in a browser-independent
way, while plugins help browsers to deal with the contents they send
and get from SPs.

Browsers' extensions, on the other hand, are a modular approach to
expand a browser's functionality in all sorts of ways. However, unlike
plugins, they are transparent to SPs. Therefore, they
are unsuited for supporting our personal services, since we want
SPs providers to be able to find out personal services of interest.

The possibility of using browsers' extensions to run our personal
services' Broker could be a possibility, but it it is preferable to
have the Broker as process independent from all browsers, since it
simplifies the configuration and the access to personal services in
a browser-independent way. Naturally, the same applies to personal
services, for which there is no added value in deploying them as
browsers' extensions.

Instead of relying in browser-dependent plugins or extensions, the
AMA (Agência para a Modernização Administrativa\footnote{Portuguese
agency that coordinates the technical innovation in the Public
Administration; it is in charge of the development and deployment of
technical solutions for handling the Portuguese EId (Electronic
Identification card, named Cartão de Cidadão) in computational
environments.}) followed a different approach for interacting with
what we call a personal service. In order to allow citizens to use
their EId (a smartcard) on any browser to authenticate themselves
in several services provided by the Public Administration, they
decided to develop an external application entitled
``\textit{plugin Autenticação.Gov}''\footnote{Autenticação.gov,
\url{https://www.autenticacao.gov.pt}}. This application, which is
exactly what we call a personal service, implements an HTTP server
that will be waiting for eID-related requests on one of the
following pre-defined TCP ports: $35153$, $43456$, $47920$, $57379$ or
$64704$. This application is invoked by JavaScript code running on
the browser.

This last approach is similar to the one already mentioned in the
introduction~\cite{Zuquete14}, which uses HTTP redirections instead
of HTTP requests from JavaScript code. The advantages of using
JavaScript are twofold: first, different ports can be used (which
tolerates collisions in port allocations), since with JavaScript it
is possible to test all pre-established ports sequentially until
finding a hit, while this is impossible with HTTP redirection codes
(because after a failure the control does not return to the SP
that did the redirection); second, a service can be implemented
together by JavaScript code and an external application. Its may
drawback, on the other hand, is that JavaScript code can make
arbitrary calls to any local HTTP services/servers, which may lead
to vulnerability exposures (e.g. The Filet-o-Firewall attack\footnote{
Filet-o-Firewall: new vulnerabilities in UPnP expose the whole network,
\url{https://www.kaspersky.com/blog/filet-o-firewall/4533}
}). Note that those vulnerabilities are smaller with HTTP
redirection codes, because target personal services must cooperate
with the redirecting service in order to return latter the control
to the attacker (with another redirection). Thus, it is impossible
for a malicious SP to conduct a request/response dialog with an HTTP
service running locally to the client browser by using HTTP
redirections.

From this last discussion it is clear that relying on JavaScript to
find and invoke personal services is to open the door to
vulnerability exploits, while doing it with requests/responses
recognized by browsers, such as HTTP redirection codes, enables
browsers to recognize intents to interact with local services and
get instructions about it from users. This is why we based our work
on HTTP redirections.

%
%
%

\section{Architecture}
\label{architecture}

The architecture that we propose to support the discovery and
invocation of personal services is based on two main components: (i)
a Broker service and (ii) a protocol to discover personal services,
using a Broker, and to invoke them.

\subsection{Broker}

The Broker is an HTTP service that keeps record of a set of
personal services. Personal services are themselves HTTP services,
i.e., they must support an ordinary HTTP interface. The Broker is
agnostic about the actual features that each personal service
provides through HTTP.

\subsubsection{Yellow- and white-pages paradigms}

The Broker supports a yellow-pages service to enable SPs to list
personal services of interest. This enables an SP to get a list of
available personal services for a given purpose, before actually
choosing one of them. The choice may even be given to the user that
is interacting with the SP through their browser. The yellow-pages
service returns a list of service descriptions that match a
particular set of attributes provided by the SP. A service
description is a set of attributes describing a personal service
(see next section).

The Broker supports a white-pages service to provide a personal
service handle to an SP. In this case, the SP must provide a set of
attributes matching a single service.

\subsubsection{Service naming paradigm}

The naming model implemented by the Broker is based on services'
attributes, similarly to~\cite{Peterson87}. This option was
motivated by the fact that, currently, plugins are associated to
specific contents by a set of attributes describing the kind of data
they are able to handle. This association between content attributes
and plugins was introduced by the NPAPI interface, created by
Netscape and used by most browsers until 2013, when the main
browsers began their phase-out.

The set of attributes associated with each personal service is not
limited. Each service can freely define the set of attributes that
may be listed or looked upon by SPs. This enables SPs and personal
services to choose the best way to find each other.

Two different sets of attributes are associated with personal
servers: \textit{configuration} and \textit{presentation}
attributes. Configuration attributes is a set of operational
attributes that is visible only to the Broker. They are meant to be
used only by the Broker for establishing a proper management of the
services. Presentation attributes, on the other hand, are visible to
SPs and are not mandated by the Broker. These are the attributes
that SPs should use to find and make use of personal services.
Therefore, in what concerns SPs, personal services are named solely
with presentation attributes.

The presentation attributes should be (short) feature descriptions, for
helping SPs to find the right personal service for a particular
purpose. The number and actual name of presentation attributes is
not limited or imposed by the Broker. For instance, content handling
personal services may have a \attrname{Purpose} attribute with the
\attrvalue{ContentHandling} value and a \attrname{ContentType}
attribute associated to an existing MIME type.  For identification
and authentication personal services, they may have a
\attrname{Purpose} attribute with the \attrvalue{Authentication}
value and an \attrname{AuthenticationType} with a description of the
supported authentication method. Note, however, that these examples
are just illustrative, they are not mandatory, other attributes may
be used. The choice of the right attributes should follow some
posterior standardization process, for enabling SPs to know how to
look for personal services of interest.

\subsubsection{Name structures}

Personal service names are formed by a set of presentation
attributes. We chose to use the popular JSON (Java Script Object
Notation~\cite{rfc7159}) format to define a set of presentation
attributes associated with each personal service. Using JSON, a
service name is an object where each member defines an attribute
string associated with a value. Values can be strings, numbers,
objects, arrays, boolean values (true/false) or no-values (null).

\subsubsection{Name listing and resolution}

An advantage that we have considered of using JSON for expressing
service names was that we could rely on existing JSON querying
syntaxes to implement both name listing and resolution processes.
However, we decided to delegate such task to SPs and
provide a minimal functionality on Brokers. In this sense, a Broker
provides a very simple attribute matching functionality for
selecting service names to be listed or resolved upon SPs' requests.

For service name listing, the Broker accepts a case-insensitive,
single attribute specification (attribute: value) for listing all
services matching this object member in their names. It does not
provide any support for regular expressions in the matching
operation. The result of a listing request is a JSON array of
service names (with all their presentation attributes) matching the
listing query.

For service name resolution, the Broker accepts a case-sensitive,
query object for resolving a service name fully matching all the
members of the query object. The query object cannot match more than
one service name. The resulting handle, returned to the browser,
will contain the complete service name matching the query.

\subsection{New HTTP status codes}

Both calls to yellow- and white-pages are made by an SP to the
browser with a new HTTP response message, i.e., a response with a
new status code (provided within the Status-Line HTTP header field).
We chose a set of status codes in the range 31X, namely $310$, $311$, $312$
and $313$, because both calls are a kind of redirection.

\begin{figure}[p]
\centerline{
\includegraphics[scale=.55]{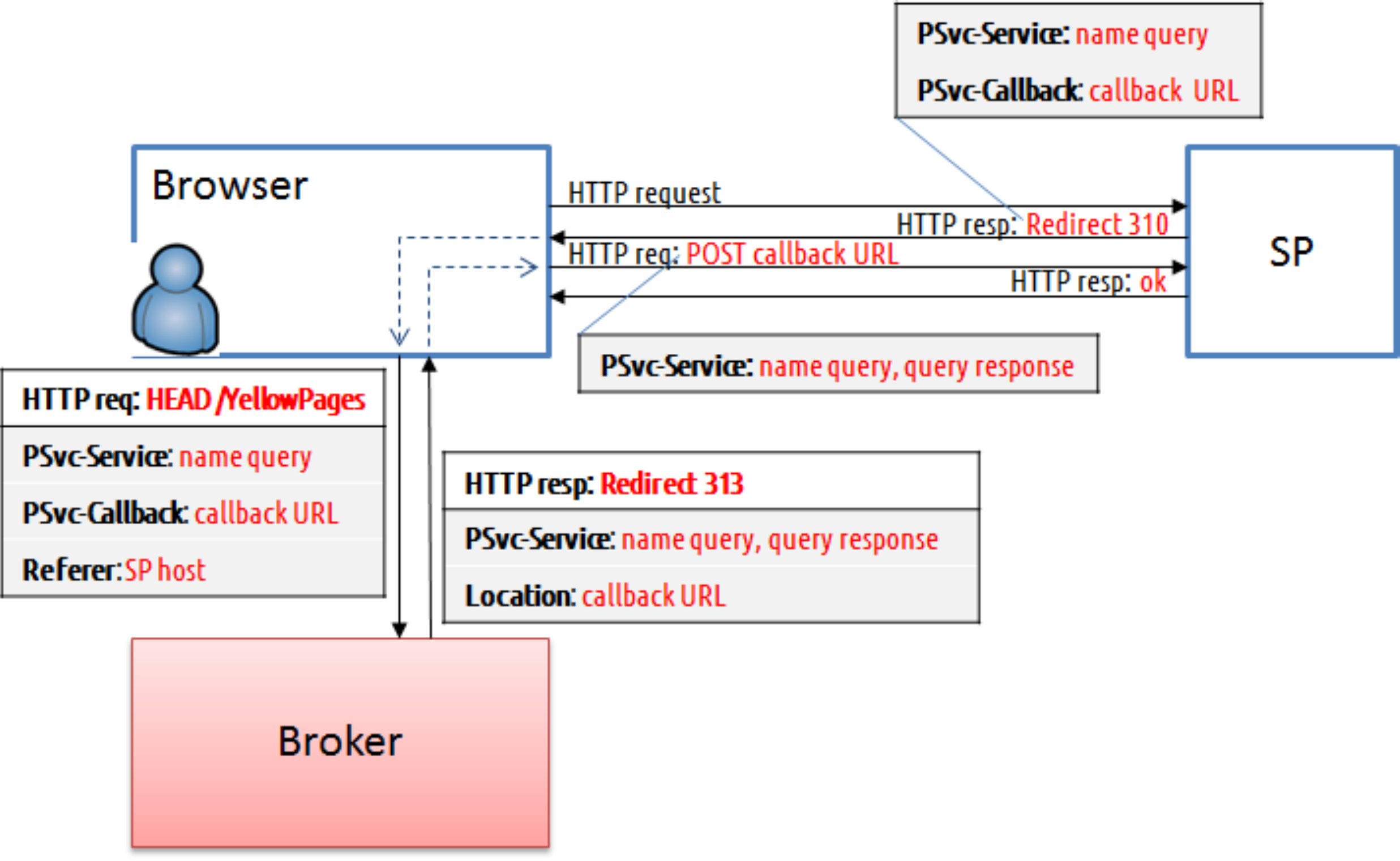}
}
\centerline{
\includegraphics[scale=.55]{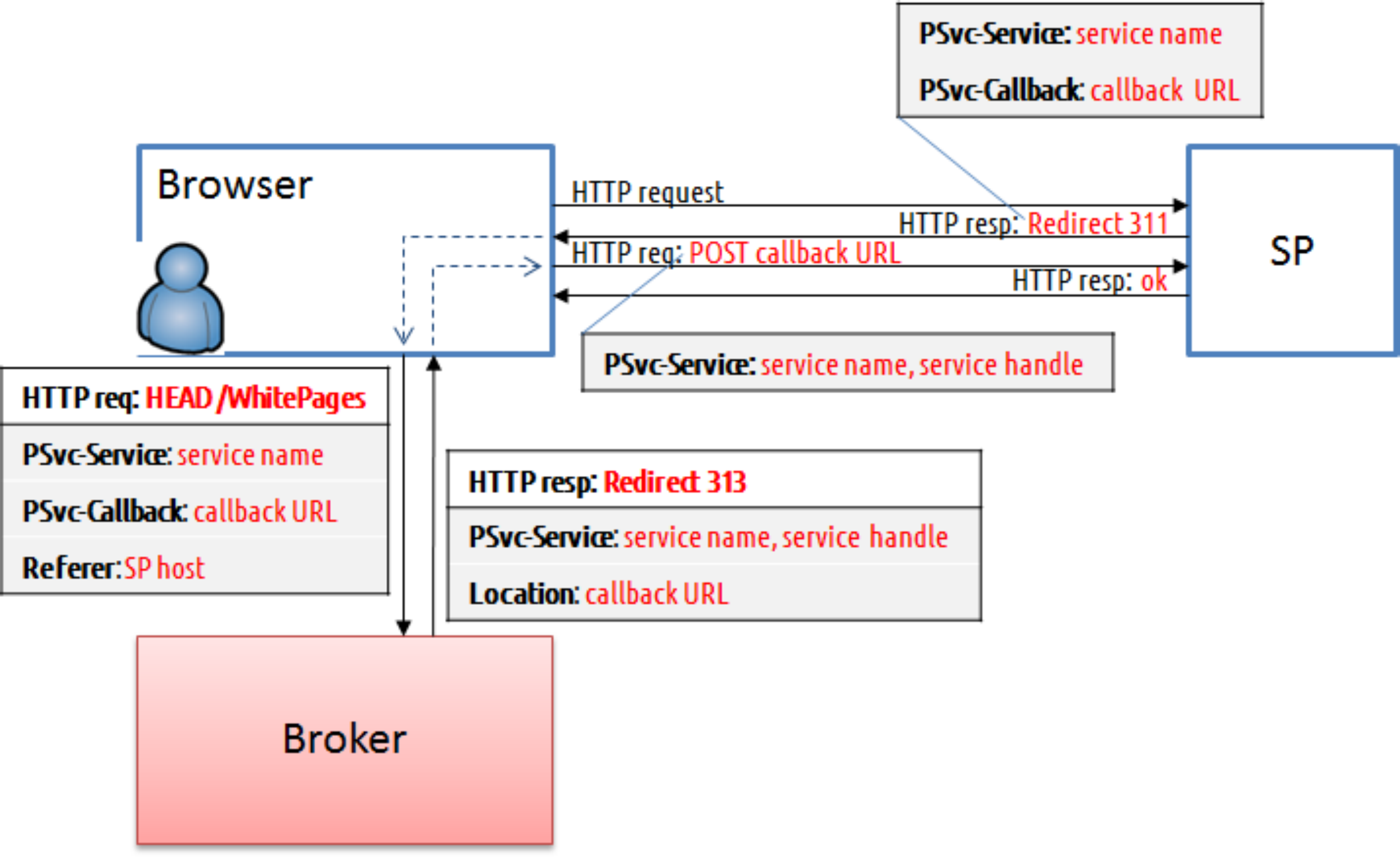}
}
\caption{Yellow-pages (service listing, top diagram) and white-pages
(name resolution, bottom diagram) services:  request/response flows
and relevant HTTP fields.}
\label{Pages.Fig}
\end{figure}

\begin{figure}[t]
\includegraphics[scale=.48]{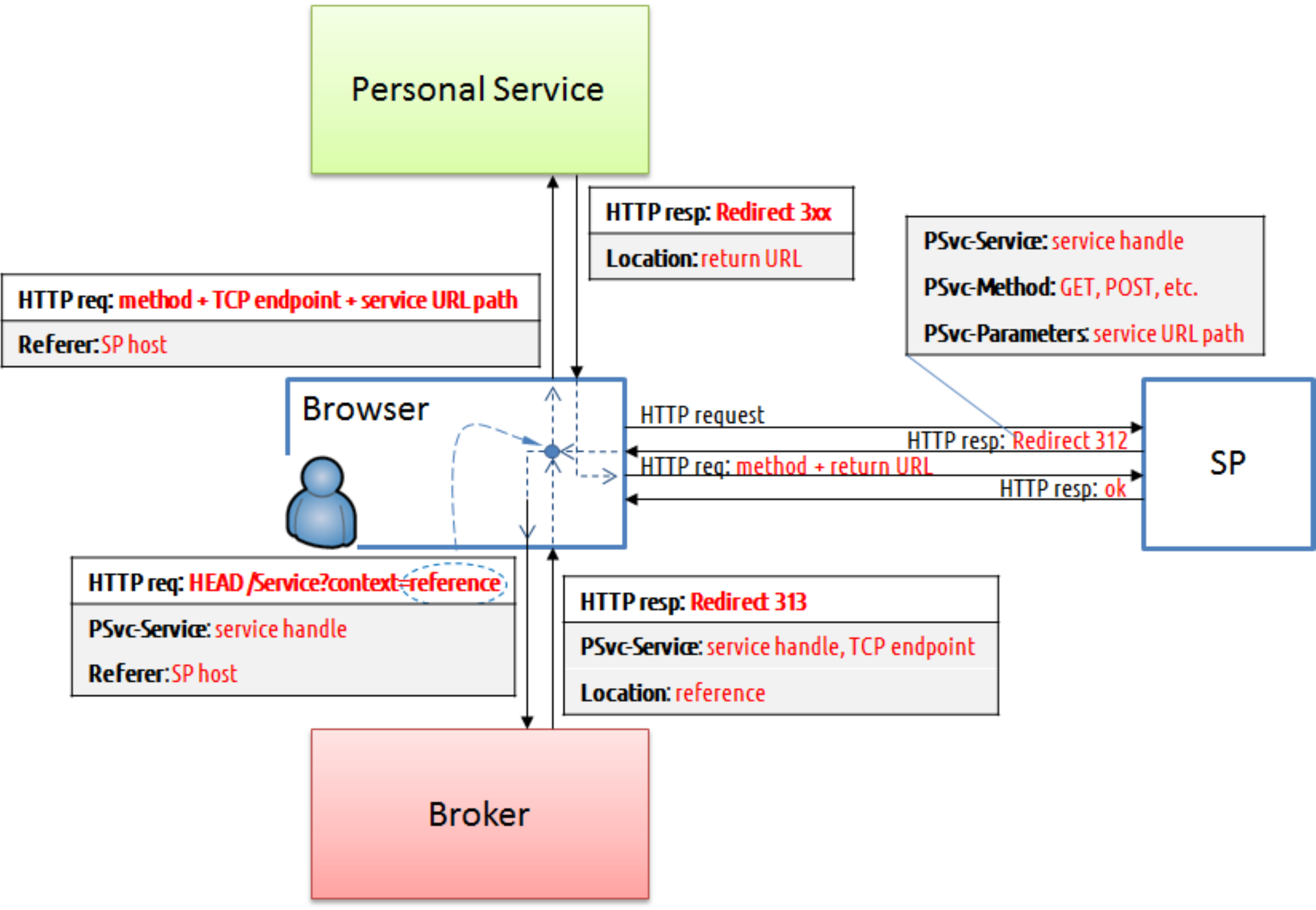}
\caption{Personal service invocation: request/response flows
and relevant HTTP fields. This diagram shows a complex example where
a personal service returns some result to an SP; this may not be
necessary for all cases, namely for content-handling personal services.}
\label{Service.Fig}
\end{figure}

The status codes $310$ to $313$ are used for the following 4
different operations (see Figures~\ref{Pages.Fig} and~\ref{Service.Fig}):
\begin{itemize}
\item {\bf 310}: Call the Broker's yellow-pages service to list personal
      services with a given attribute (Figure~\ref{Pages.Fig}, top).
\item {\bf 311}: Call the Broker's white-pages service to resolve a
      personal service name (set of attributes given by a query
      object) into a handle (Figure~\ref{Pages.Fig}, bottom).
\item {\bf 312}: Call a personal service given a handle to it (see
      Figure~\ref{Service.Fig}).
\item {\bf 313}: Redirect a result provided by the Broker to any of
      the previous calls.
\end{itemize}

\noindent
A set of extra header fields in the HTTP response message are used to
parametrize an SP request. We defined the following ones:
\begin{itemize}
\item{\texttt{PSvc-Service}} (mandatory): the value of this
      field is a JSON object with key-value pairs. For the
      yellow-pages service, it contains the attribute of the services to be
      listed. For the white-pages service, it contains the
      attributes of a service to get an handle to. Finally, for
      invoking a personal service it contains the handle for that
      service.
\item{\texttt{PSvc-Method}} (optional): its value is the method to be used
      in an HTTP request when invoking a personal service.
      It may be obtained from the service's presentation
      attributes. When not specified, it defaults to GET.
\item{\texttt{PSvc-Parameters}} (optional): its value is a path plus
      query to be used in an HTTP request to invoke the personal service.
      The totality or only part of its value may be obtained from the
      service's presentation attributes.
\item{\texttt{PSvc-Callback}} (mandatory): its value is an URL to be
      used in HTTP requests to return the control back to the SP
      (upon a call to the Broker).
\end{itemize}

\subsubsection{Handling status code 310-312 redirections}

When a browser receives a status code $310$ or $311$ on a response
message (from an SP), it transforms it into a HEAD request to the
Broker. The normal reply to these messages is provided to the
browser with a status code $313$, which implies a further
redirection to the SP with the Broker's result. The redirection is
made to a callback URL that must be provided in the
\hfield{PSvc-Callback} header field.

When a browser receives a status code $312$ on a response message
(from an SP), it transforms it in a request to a personal service.
Such transformation involves first a translation, by the Broker, of the given
service handle to its TCP endpoint (hostname and port, see
Figure~\ref{Service.Fig}). This endpoint is then combined with the
value of the \hfield{PSvc-Parameters} field to build the URL to
invoke the personal service. Finally, all header fields existing in
the SP response, plus its body, are copied to the personal service
request.

The reply to such request is provided by the target personal
service, which may call back the SP, if necessary, using the already
existing HTTP redirection mechanisms and a callback provided by the
SP (specified with a mechanism other than the \hfield{PSvc-Callback}
header field). Note that if redirections are used to return the
control from the personal service to the SP, the
method (GET, POST, etc.) used to invoke the personal service is the same that will be
used to call back the SP. This potential limitation can be overcome
by returning the control to the SP from a Web page returned by the
personal service, using or not JavaScript (see
Section~\ref{experimentation}).

The HEAD messages used to invoke the yellow- and white-pages
functions of the Broker are similar to GET messages, but their
response has no body. Thus, Broker responses to SP calls are
conveyed to the calling browser in the header field
\hfield{PSvc-Service}. The redirection of
the response from the browser to the SP depends on the service
requested to the Broker:
\begin{itemize}
\item{\bf Yellow-pages}. The reply contains the name query and a list of service
     descriptions that should be uploaded to the SP and possibly
     presented to the user through their browser. In this sense,
     the response to the SP should be made by a POST request, which
     is triggered by the Broker by replying with a $313$ redirection
     to the SP's callback URL (given by the \hfield{Location} header field).

     The list of service descriptions, given by the
     \hfield{PSvc-Service} header field, will be a
     JSON array of the objects with the presentation attributes of
     those services.
\begin{lstlisting}[basicstyle=\ttfamily\small,frame=single]
PSvc-Service = {
  "operation" : "Yellow Pages",
  "request" : { attr_name : attr_value },
  "response" : [ services' presentation attributes ]
}
\end{lstlisting}

\item{\bf White-pages}. The reply contains a personal service handle
     that should be uploaded to the SP. Once reaching the SP, can be
     invoked immediately or later on, for instance, upon a user
     confirmation. As in the previous case, the response to the SP
     should be made by a POST request, which is triggered by the
     Broker by replying with a $313$ redirection to the SP's callback URL.

     The service handle and its description, also given by the
     header field \hfield{PSvc-Service},
     will be a JSON object with two elements: (i) the service's
     presentation attributes and (ii) its handle.

\begin{lstlisting}[basicstyle=\ttfamily\small,frame=single]
PSvc-Service = {
  "operation" : "White Pages",
  "request" : { service attributes },
  "response" : {
    "service" : { service presentation attributes },
    "handle" : service handle
  }
}
\end{lstlisting}

     The handle is an opaque datum that only makes sense for
     the Broker that produced it. In other words, neither browsers
     nor SPs have to interpret the contents of handles.
\end{itemize}

\subsubsection{Brokers' deployment}

Different Broker deployment strategies can be considered. For
instance, browsers can use their own configuration to manage their
Brokers. In this case, they could use different Brokers, and Brokers
could be part of browsers' distribution packages, instead of
independent applications. Different Brokers could use or not the
same configuration files for the personal services.

This last approach does not create any technical problems to exploit
personal services but it has the potential to introduce entropy in
the users' execution environments. In fact, it would be much simpler for
a user to have a single Broker, independent from all browsers, with
a well-known communication end-point, that a user could recognise
and get used to, to manage the exploitation of their personal
services.

Consequently, the architecture that we uphold is the following:
\begin{itemize}
\item A single, per-user repository of personal services' definitions.
      This repository can be a set of JSON-formated text files, editable with
      ordinary, text editing tools.
\item A single, per-user Broker application. This application may be
      launched by any browser if not already running. The TCP endpoint
      used by the Broker should not be unique, to allow other Brokers
      to coexist, namely the ones of other users in multi-user
      systems.

      In Unix-like systems one can use TCP over Unix sockets, which
      can be given a name under the user's home directory file
      system hierarchy, to easily avoid clashes between endpoint
      names given to different Brokers. In Windows systems we need
      to use different TCP ports for a single loopback address or
      different loopback addresses.
\end{itemize}

\subsection{Management of handles}
\label{handles}

%

Handles are opaque data structures for browsers, in the sense that
they should not interpret their intrinsic value. The value of
handles is defined by Brokers and should be used only by them to
manage personal services. When a browser receives a handle-based
service call, it first calls the Broker to get the current TCP
endpoint name for the service corresponding to the handle (see
Figure~\ref{Service.Fig}).

In the case the handle refers a local personal service, the Broker
will launch it, if not already running; at the end, it redirects the
browser to the correct TCP endpoint using again the $313$ status
code but with a \hfield{Location} field with an reference to an
internal, ongoing call to the personal service. This internal
reference is signaled (i.e., differentiated from a regular URL) by a
location string formed by a colon followed by a value with the
internal reference, a value provided by the browser in the
parameters of the HEAD request.

In the case the handle refers to a remote personal service, the
Broker checks if the remote personal service is accessible and if
so, it proceeds as if it were a local service.

In Figure~\ref{Service.Fig} we can observe the complete sequence of
requests and responses involving the SP, the browser, the Broker and
the personal service. In this sequence the browser keeps a service
request on hold, while waiting for a Broker's response. Upon such
response, the browser creates an HTTP request to the service
using parameters extracted from both the SP response and the Broker
response. The method used in the request is given by the SP in the
\hfield{PSvc-Method} header field (or GET by default). Although not
shown in the figure, the browser copies all the headers of the SP
response to the personal service request, this way allowing a more
flexible interaction between them.

\subsection{Access control policies and mechanisms}
\label{access.control}

Currently, browsers allow SPs to invoke HTTP services running in a
browser's host; all they have to do is to explore HTTP redirections
to an TCP/IP address locally bound to the host itself (the
``localhost'' name, or an IP address {\localhost}, and some TCP
port). As previously referred at the end of
Section~\ref{http.redirections}, this was explored
in~\cite{Zuquete14} to explore a single personal service.

An HTTP redirection to the localhost may look suspicious, however
most browsers cope with it without warnings. From our experience,
only Opera warns the user and asks for permission to proceed. In
our case, the access to personal services requires service handles,
and these can only be obtained upon a service name resolution. Note
that personal services can rule out accesses using the previously
referred redirection to localhost (i.e., with status codes $3$xx
other than the $312$) by checking the presence of the special HTTP
headers included by the browser upon a handle-based service
invocation (not shown in Figure~\ref{Service.Fig} for simplicity,
but referred at the end of section~\ref{handles}).

Therefore, two architectural elements can implement access
control policies regarding the access of SPs to personal services:
\begin{itemize}
\item Broker: it can allow (whitelisting) or prevent (blacklisting)
      some services from being listed or resolved by certain SPs.
\item Browser: it can allow a handle to be used by any SP, allowing
      related services to share a handle, or restricting the
      exploitation of a handle only to the SP that received it from
      the browser's Broker.
\end{itemize}

Regarding the first case, the implementation of an access control
policy when dealing with service names, it does not need to be
universally defined, it can be freely implemented by Brokers. In
fact, users are free to pick any suitable Broker, and different
Brokers can implement different protection mechanisms and allow
different protection policies, while keeping a standard communication
protocol with local browsers and SPs.

Regarding the second case, it needs clarification because all
browsers should follow a similar strategy. Furthermore, browsers
should cope with their Brokers to provide a clean and well-defined
access control policy. The easiest way to deal with such cooperation
is to leave all access control decisions to Brokers, being them the
sole entities responsible for taking care of such decisions. This
can easily be implemented, since a handle-based call to a personal
service needs to be resolved by the Broker to a TCP endpoint, and
for that purpose the Broker receives the identity of the SP host
(together with the service handle, see Figure~\ref{Service.Fig}).

\subsection{Content transfer in service calls}

When SPs call personal services, using the $312$ redirection, they
can transfer arbitrary data to services by adding it to the reply
header and body. With the exception of the \hfield{Status} header
field (always the first header field), the entire header and body
are copied to the request that is created to invoke the
personal service. The only header field that is added in this
response-to-request transformation is the \hfield{Referer} one,
identifying the SP (see Figure~\ref{Service.Fig}). Note that this is
not what happens with the actual redirections, where both the header
and the body of the response are ignored.

The same thus not happen when the personal service returns the
control to the SP, because this is done with the actual
redirections. However, the personal service is free to interact
directly with the SP, without involving the browser in that
dialog.


\subsection{Security-related limitations}

The redirection $313$ was conceived to implement Brokers' responses
to calls triggered by redirections $310$, $311$ or $312$. 
Therefore, browsers should restrict the use of redirections $313$
only to their Broker, and deny their arbitrary use as other $3$xx
redirections orchestrated by SPs. Otherwise, they could be used to
implement new attack vectors, where malicious SPs could force
browsers to make unintended requests to victim servers.

\subsection{Error handling}

When wrong HTTP redirections occur, they result in HTTP access errors
presented by browsers to users; no feedback is given to the agent
that gave the redirection order. In our case, however, we can and
should provide some error reports to SPs upon some errors that can
occur while exploiting our special redirection mechanism.

We have identified the following relevant error situations, some of
which should trigger well-defined error reporting interactions:
\begin{itemize}
\item The browser is not prepared to deal with personal services.
      In this case, SPs must be warned in advance if browsers
      support personal services, because the opposite is impossible.
      This capability can be signaled by a special field in all
      HTTP requests made by the browser:
      \hfield{PSvc-Version}. This field must contain the
      list of protocol versions supported by the browser to interact
      with personal services.
\item The Broker cannot be called for listing services or resolving
      service names. The browser should report this error to its
      user and should carry on with the reply to the SP with an
      empty result.
\item Several other errors may happen upon an SP call to the Broker
      or to a personal service. This case is more delicate than the
      former, since there is no standard response to be provided to
      the SP by either the browser or the Broker.
      We decided to use the SP callback path provided by the
      SP (in the \hfield{PSvc-Callback} header field), to carry on
      an error reporting call. In such call we use an header field,
      \hfield{PSvc-Error}, to convey four possible textual error
      codes:
      \begin{itemize}
      \item ``\texttt{parameters}'': when the SP request is malformed (has missing or wrong parameters).
      \item ``\texttt{ambiguous}'': when more than one service matches a White Pages' request.
      \item ``\texttt{handle}'': when the service handle provided is no longer valid.
      \item ``\texttt{service}'': when the requested service cannot be found or activated.
      \end{itemize}

      For interactions between SPs and personal services not
      requiring a posterior SP callback, SPs can still handle this
      error by providing a callback path just for this purpose.
\end{itemize}

\section{Implementation}
\label{implementation}

We built a prototype For testing the previously described
redirections and name service. In this prototype we did not change any
browser; instead, we implemented an HTTP proxy which deals with the
new redirections. This is not a suitable solution for all cases, since
it cannot deal with redirections when they occur withing HTTPS
sessions. However, it facilitated a lot the creation of a
proof-of-concept implementation.

The core of the prototype is then an HTTP proxy and a Broker. Both
where implemented in Java, using a set of classes and interfaces
belonging to packages under \texttt{com.sun.net}:
\texttt{HttpServer} for implementing HTTP servers,
\texttt{HttpHandler} to define service handlers and
\texttt{HttpExchange} for implementing HTTP interactions.

The proxy forwards GET and POST requests, adding to them the
\hfield{PSvc-Version} header field. The subsequent responses of SPs
are first analyzed, to handle the new redirections. If not present,
the response is forwarded to the client browser; otherwise, a new
response is created by calling a Broker.

\subsection{Configuration files}

The interaction between the HTTP proxy and the Broker is helped by a
file (\texttt{broker.ept}) stored in the \texttt{.PS} subdirectory
of the user's home directory. This file is the Broker's contact
endpoint; in Windows systems it is a regular file containing its TCP
port (the IP address is implicitly \texttt{127.0.0.1}); in Unix-like
systems, it is a UNIX socket name.

For each personal service there should be a separate file in the
\texttt{.PS} directory. We used the extension \texttt{\.psd} (from
personal service description) to identify such files from the
Broker. Each of these files should contain a single object
description, in JSON, with two sub-objects: \texttt{configuration}
and \texttt{presentation} (for the configuration and presentation
attributes, respectively). The contents of the \texttt{presentation}
object are free, while the \texttt{configuration} object should
contain the following attributes that are looked for by the Broker:
\begin{itemize}
\item \texttt{dir}: the directory where the service should
      run; if absent, the Broker uses \texttt{.PS}.
\item \texttt{cmd}: the command that starts the personal service,
      if not yet running.
\end{itemize}
Those attributes allow the Broker to act similarly to the well-known
\texttt{inetd} Internet service daemon (or super-server).

For completeness, there is also a \texttt{broker.psd} file in the
\texttt{.PS} directory with the configuration attributes for the
Broker. It allows our HTTP proxy (or a modified browser) to launch
the Broker if not yet running.

\subsection{Interaction with personal services}

The TCP port used by each personal service is allocated by the
Broker and passed as a parameter when launching the personal
service. The Broker allocates it for all local addresses and sets
its to be reused by another process; the personal service binds it
only to a local address (\texttt{127.x.x.x}).

Upon launching a personal service, the Broker keeps a local record
of its process and TCP endpoint within its internal structure
associated with the service. This structure is used each time an SP
invokes the personal service using a handle to it. If the service
is still alive, its TCP endpoint is returned to the Broker caller;
otherwise, the personal service is launched again, possibly using a
different TCP endpoint. This way, we are able to accommodate both
personal services that terminate after completing a service request
and those that stay alive after a first request.

In our implementation, a handle is a JSON object with the
description of the service that requested it and a reference to the
internal structure describing the service, encrypted with a
symmetric key known only by the broker. This way, SPs cannot
generate valid handles and cannot transfer handles among themselves;
they need to go through the client's White Pages service for being
subject to authorization clearance.

\section{Experimentation}
\label{experimentation}

For experimenting the invocation of personal services we adapted the
eID-based authentication protocol presented in~\cite{Zuquete14}. The
adaptation consisted in the following actions:
\begin{itemize}
\item The SP requesting a user authentication checks if the client
      supports the invocation of personal services (by checking the
      presence of the \hfield{PScv-Version} field).
\item Upon a successful validation, the SP check if there is a
      service capable of performing the eID-based
      authentication protocol described in~\cite{Zuquete14}. For
      this, it uses the White Pages' service to lookup for a service
      with the following attributes:
\begin{lstlisting}[basicstyle=\ttfamily\small,frame=single]
{
  "Purpose": "authentication",
  "Device": "Portuguese eID"
}
\end{lstlisting}
\item If there is such a service, the SP invokes it with the
      redirection $312$ using a GET method.
\end{itemize}

The existing personal service was not changed except for using the
TCP port provided as parameter to wait for HTTP requests. Its
configuration file, used by the Broker, was created with the
following contents (for a Windows system deployment):

\begin{center}
\begin{minipage}{\linewidth}
\begin{lstlisting}[
    basicstyle=\ttfamily\small,frame=single,
    extendedchars=true,
    literate={ã}{{\~a}}1]
{
  "configuration" : {
    "dir": "Z:/PersonalServices/CCPersonalService",
    "cmd": [
      "java",
      "-jar",
      "CCPersonalService.jar"
    ]
  },
  "presentation": {
    "Purpose": "authentication",
    "Credentials": "digital signature",
    "Protocol": "certificate + digital signature",
    "Device": "Portuguese eID",
    "Device name": "Cartão de Cidadão"
  }
}
\end{lstlisting}
\end{minipage}
\end{center}


The experiment ran without any problems. The personal service
handling the eID authentication, yet another Java application, was
launched upon the first SP call and remained executing, tackling
posterior requests. Comparing with the original protocol design, the
SP no longer has to know the exact port where the personal service
of interest is waiting for requests.

\begin{figure}[ht]
\includegraphics[width=\textwidth]{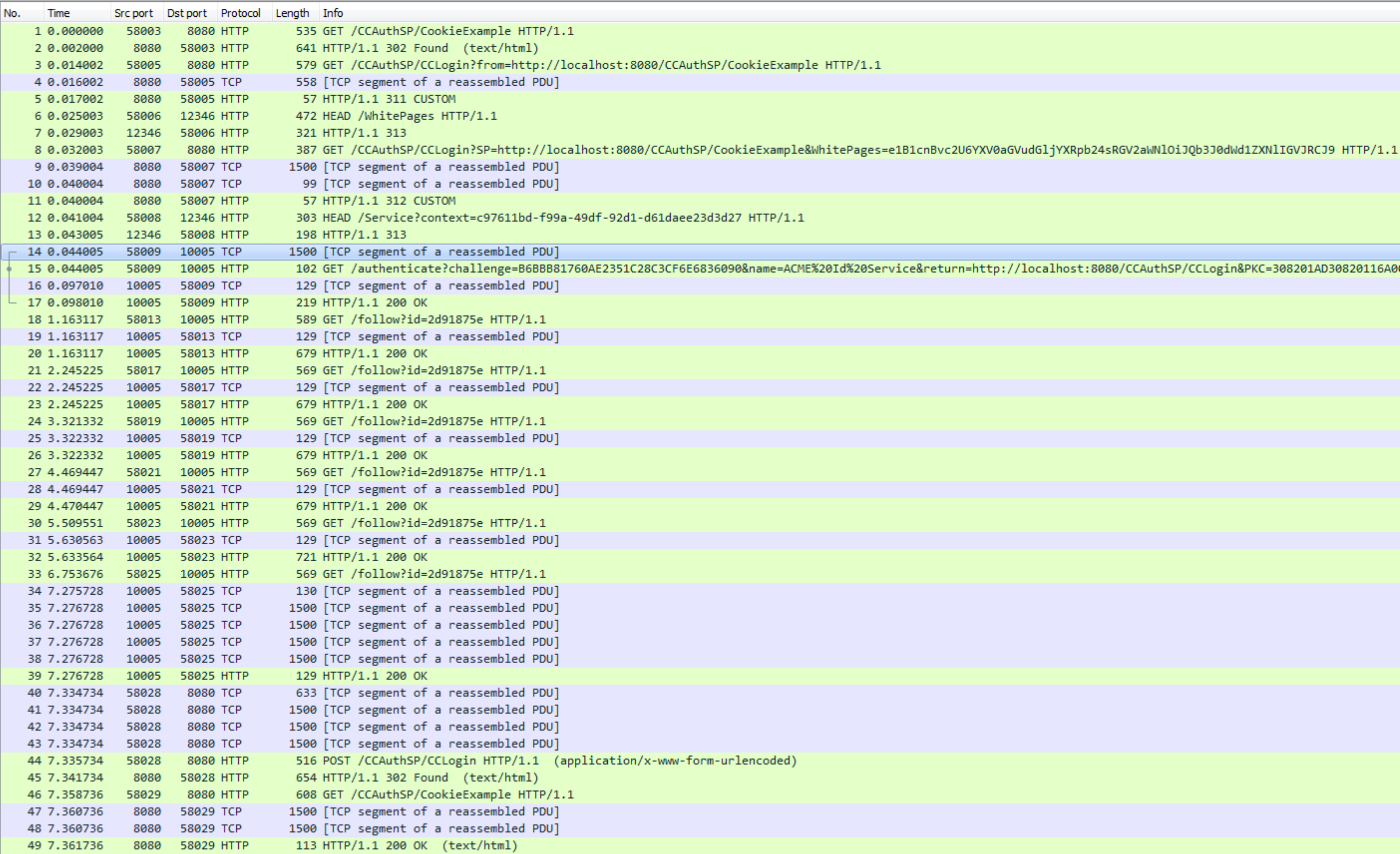}
\caption{Capture of a dialog between a browser (its proxy,
actually), an SP, a Broker and a personal service. The SP uses
port 8080, the Broker uses port 12346 and the personal service uses
port 10005. The invocation of the White Pages service can be
observed in messages 5 to 8, while the invocation of the personal
service using a previously obtained handle can be observed in
messages 9 to 44.}
\label{capture.Fig}
\end{figure}

Figure~\ref{capture.Fig} shows the packet capture of the entire
interaction involving the proxy, an SP, the Broker and the involved
personal service. We left the browser out because it would not highlight
any special aspect of the message exchange. The SP uses port 8080,
the Broker uses port 12346 and the personal service uses port 10005.

The SP implements a demo service that manages client cookies, namely
authentication cookies, created upon a successful eID-based
authentication. In message 1 the server receives the initial
request, which does not has any cookies, which it redirects to its
own eID-based authentication service (using redirection 302, message
2).  This service, invoked in message 3, starts by using the
client's White Pages to check whether it has the appropriate
personal service (messages 4 and 5).  The invocation of the White
Pages service is visible in messages 6 and 7, and the result is
conveyed to the SP in message 8. Once having a proper handle, the SP
uses it to invoke the personal service of interest (messages 9, 10
and 11). Upon that, the proxy uses the handle to get the TCP
endpoint of the personal service (messages 12 and 13) and the
service is invoked in messages 14 and 15 (using a GET method,
specified by the SP). Thereafter, and until message 39, the personal
service conducts the eID-based authentication while keeping a dialog
with the user through the browser. This dialog terminates upon a
successful eID authentication, which return an HTML content that
performs an automatic POST of the authentication results to the SP
(messages 40 to 44). Upon the validation of the results, the SP
redirects the browser to the initial service called by the user,
using again the redirection 302 (messages 45 to 49).

This experience does not show the invocation of the Yellow Pages
service, because it was not necessary in this demonstration.

\section{Access to Brokers by scripts of Web resources}
\label{javascript}

As already mentioned in Section~\ref{related.work}, a solution that
is nowadays in place to invoke a personal service that deals with
the eID-based authentication using the Portuguese identity card
(Cartão de Cidadão) is based on the use of a list of fixed TCP
endpoints and active contents (JavaScript scripts) that look for
them. This solution poses an interesting question: should browsers
allow JavaScript, or any other scripting language interpreted within
the context of a Web resource, to invoke the browser's Broker?

With the proposed model for providing responses to invocations to
the Broker, that would not be possible. In fact, the proposed model
uses the redirection 313 to return the control to an SP (a server), and
not to a JavaScript client. Furthermore, the identification of the
requester of the Broker's services (provided in the
\hfield{Referrer} header field) would have to be extracted from the
DOM (Document Object Model) of the Web resource executing the
script, in order to provide an accurate access control decision by
the Broker (as discussed in Section~\ref{access.control}).

Concluding, this is an interesting topic for future work and discussion.

\section{Conclusions}
\label{conclusions}

In this paper we presented an architecture to allow SPs to interact
with personal services of interest. Personal services are HTTP
services that run under the control of the browser user, either on
the same machine of the browser or in a different machine, and
provide some service that may be useful for the SP. The example that
we have used, throughout this paper, was a personal service
developed for performing an eID-based authentication.

For enabling SPs to make request to personal services we developed a
name service to help in that task. The name service allows SPs to
discover personal services of interest, to get an handle for a
particular personal service and to call that personal service using
that handle. The name service is supported by a Broker, which is an
application that runs and exists independently of browsers. The
Broker is responsible for knowing which personal services exist, for
running them and for managing their communication endpoint.

The interaction between SPs, browsers and the Broker is mainly
supported by a new set of $3$xx redirections, complemented with some
new HTTP header fields. The interaction includes the invocation of
Yellow and White pages on the Broker and a handle-based call to a
personal service.  

The proposed architecture was implemented on an HTTP proxy and successfully
tested with an adapted implementation of the authentication protocol
described in~\cite{Zuquete14}.

\bibliographystyle{plain}
\bibliography{main,rfc-index}

\end{document}